\documentclass[aps,preprint]{revtex4}

\usepackage{amssymb,amsmath,graphicx,times}

\begin{document}

\title{Impact of memory on human dynamics}

\author{Alexei Vazquez\\
Center for Cancer Systems Biology, Dana Farber Cancer Institute\\
44 Binney St, Boston, MA 02115, USA\\
Department of Physics and Center for Complex Networks Research\\
University of Notre Dame, IN 46556, USA}

\date{\today}

\begin{abstract}

Our experience of web access slowing down is a consequence of the
aggregated web access pattern of web users. This is just one example among
several human oriented services which are strongly affected by human
activity patterns. Recent empirical evidence is indicating that human
activity patterns are characterized by power law distributions of
inter-event times, where large fluctuations rather than regularity is the
common case. I show that this temporal heterogeneity can be explained by
two mechanisms: (i) humans have some perception of their past activity
rate and (ii) based on that they react by accelerating or reducing their
activity rate. Using these two mechanisms I explain the inter-event time
statistics of Darwin's and Einstein's correspondence and the email
activity within an university environment. Moreover, they are typical
examples of the the accelerating and reducing class, respectively. These
results are relevant to the system design of human oriented services.

\end{abstract}

\maketitle

\bibliographystyle{apsrev}

Human activity patterns are inherently stochastic at the single individual
level. Understanding this dynamics is crucial to design efficient systems
dealing with the aggregated activity of several humans.  A typical example
is a call center design, where we save resources by taking into account
that all workers will no call or receive calls at the same time
\cite{phone-design,reynolds03}. There are several other examples including
the design of communication networks in general, web servers, road systems
and strategies to halt epidemic outbreaks \cite{eubank04,goldenberg05}.

The stochasticity present in the human dynamics has been in general
modeled by a Poisson processes characterized by a constant rate of
activity execution \cite{phone-design,reynolds03,eubank04}.  
Generalizations to non-stationary Poisson processes has also been
considered taking into account the effects of seasonality
\cite{hidalgo06}. Yet, these approaches fail when confronted with recent
empirical data for the inter-event time statistics of different human
activities \cite{barabasi05,dezso05,oliveira05,vazquez06}. I show that the
missing mechanism is a key human attribute, memory.

\section{The model}

Consider an individual and an specific activity in which he/she is
frequently involved, such as sending emails. The chance that the
individual execute that activity (event) at a given time depends on the
previous activity history. More precisely, (i) humans have a perception
of their past activity rate and (ii)  based on that they react by
accelerating or reducing their activity rate. Although it is obvious that
we remember what we have done it is more difficult to quantify this
perception. In a first approximation I assume that the perception of our
past activity is given by the mean activity rate. I also assume that based
on this perception we then decide to accelerate or reduce our activity
rate. In mathematical terms this means that if $\lambda(t)dt$ is the
probability that the individual performs the activity between time $t$ and
$t+dt$ then

\begin{equation}
\lambda(t) = a \frac{1}{t}\int_0^t dt^\prime \lambda(t^\prime)\ ,
\label{ll}
\end{equation}

\noindent where the parameter $a>0$ controls the degree and type of
reaction to the past perception. When $a=1$ we obtain
$\lambda(t)=\lambda(0)$ and the process is stationary. On the other hand,
when $a\neq0$ the process is non-stationary with acceleration ($a>0)$ or
reduction ($a<1$).

Implicitly in (\ref{ll}) is the assumption of an starting time ($t=0$).  
For the case of daily activities this can be taken as the time we wake up
or we arrive to work. More generally it is a reflection of our bounded
memory, meaning that we do not remember or do not consider relevant what
took place before that time. For instance, we usually check for new emails
every day after arriving at work no matter what we did the day before.

Equation (\ref{ll}) can be solved for any $a$ resulting in

\begin{equation}
\lambda(t) = \lambda_0 a \left(\frac{t}{T}\right)^{a-1}\ ,
\label{lambda}
\end{equation}

\noindent where $\lambda_0$ is the mean number of events in the time
period under consideration $T$. Due to the stochastic nature of this
process the inter-event time $X$ between the two consecutive task
executions is a random variable. We denote by $F(\tau)= {\rm Prob}\left(
X<\tau \right)$ and $f(\tau)=\dot{F}(\tau)$ the inter-event distribution
and probability density function, respectively. Within short time
intervals $\lambda(t)$ is approximately constant and the dynamics follows
a Poisson process characterized by an exponential distribution
$F(\tau,\lambda(t))=1-e^{-\lambda(t)\tau}$. Furthermore, the mean fraction
of events taking place within this short time interval is
$\lambda(t)dt/\lambda_0 T$. Integrating over the whole time period we
finally obtain

\begin{equation}
F(\tau) =
\int_0^T dt \frac{\lambda(t)}{\lambda_0 T}
\left(1-e^{-\lambda(t)\tau}\right)\ .
\label{Ftau}
\end{equation}

\noindent For the stationary process ($a=1$) we recover the exponential
distribution $F(\tau)=1-e^{-\lambda_0\tau}$ characteristic of a Poisson
process. More generally, substituting (\ref{lambda}) into (\ref{Ftau}) we
obtain

\begin{equation}
F(\tau) = \left\{
\begin{array}{ll}
\displaystyle
1 - \exp\left( - \frac{\tau}{\tau_0} \right) +
\left( \frac{\tau}{\tau_0} \right)^{\frac{a}{1-a}}
\Gamma\left(\frac{1-2a}{1-a},\frac{\tau}{\tau_0} \right)
\ , & a<1\\
\displaystyle
1 - e^{ -\lambda_0\tau}\ , & a=1\\
\displaystyle
1 - \exp\left( - \frac{\tau}{\tau_0} \right) +
\left( \frac{\tau}{\tau_0} \right)^{-\frac{a}{a-1}}
\left[ \Gamma\left(\frac{2a-1}{a-1}\right) - 
\Gamma\left(\frac{2a-1}{a-1},\frac{\tau}{\tau_0} \right) \right]
\ , & a>1
\end{array}
\right.
\label{Ftaua}
\end{equation}

\noindent where $0\leq\tau\leq T$, $\Gamma(\beta,y)=\int_y^\infty dx 
e^{-x}x^{\beta-1}$ is the incomplete gamma function and 

\begin{equation}
\tau_0 = \frac{1}{a\lambda_0}
\label{tau0}
\end{equation}

\noindent for all $a\neq0$.

$a>1$: In the acceleration regime the probability density function exhibits
the power law behavior

\begin{equation}
f(\tau) = \frac{1}{\tau_0}
\frac{a}{a-1} \Gamma\left( \frac{2a-1}{a-1} \right)
\left( \frac{\tau}{\tau_0} \right)^{-\alpha}
\ ,
\label{ftauaoo}
\end{equation}

\noindent for $\tau_0\ll\tau<T$, where

\begin{equation}
\alpha = 2 + \frac{1}{a-1}\ .
\label{alphaaoo}
\end{equation}

\noindent This approximation is valid provided that $\tau_0\ll T$, i.e.
when a large number of events is registered in the period $T$.

$1/2<a<1$: In this case $f(\tau)$ does not exhibit any power law behavior. 

$0<a<1/2$: In the reduction regime the probability density function also
exhibits a power law behavior

\begin{equation}
f(\tau) = \frac{1}{\tau_0}
\frac{a}{1-a} \Gamma\left( \frac{1-2a}{1-a} \right)
\left( \frac{\tau}{\tau_0} \right)^{-\alpha}
\ ,
\label{ftaua0}
\end{equation}

\noindent but in the range $\tau\ll\tau_0$ and with exponent

\begin{equation}
\alpha = 1 - \frac{a}{1-a}\ .
\label{alphaa0}
\end{equation}

\noindent This approximation is particularly good for $\tau_0\gg T$, i.e.
when a small number of events is registered in the period $T$.

\section{Comparison with empirical data}

To check the validity of our predictions we analyze the regular mail
correspondence of Darwin and Einstein \cite{oliveira05} and an email
dataset containing the email exchange among 3,188 users in an university
environment for a period of three months \cite{eckmann04}.

\begin{figure}[t]
\centerline{\includegraphics[width=3.8in]{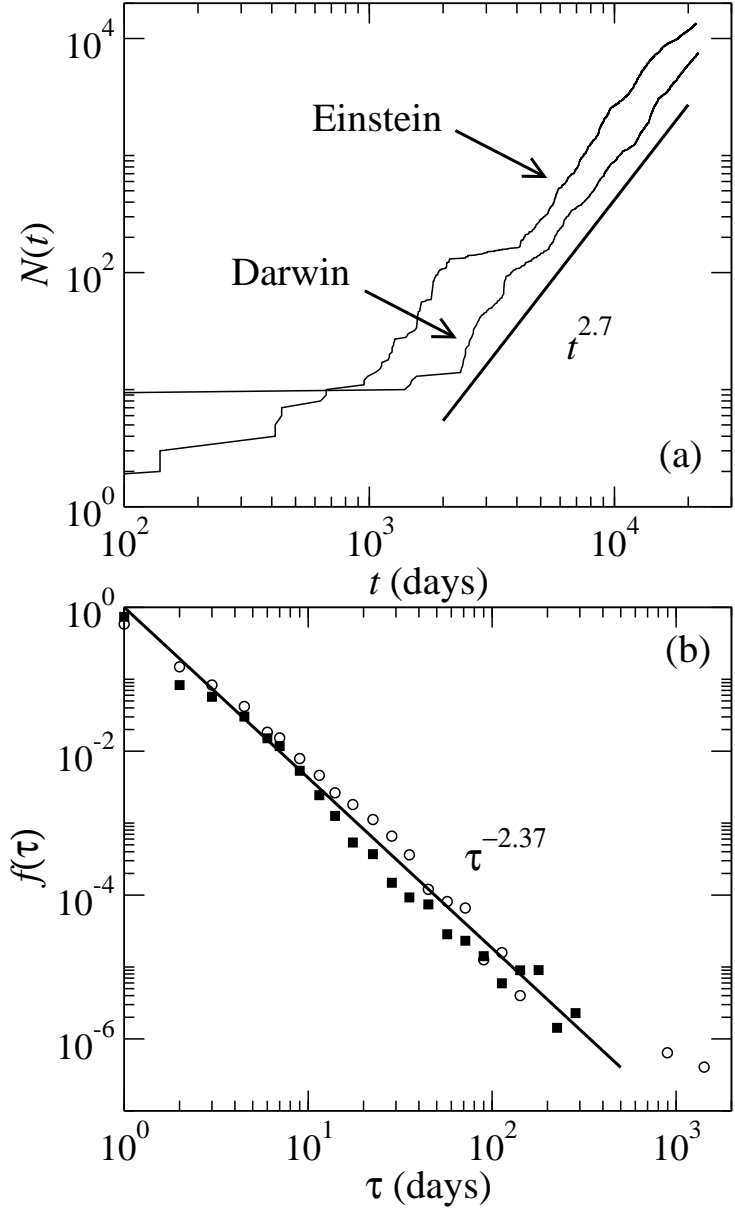}}

\caption{{\bf Regular mail activity:} Statistical properties of the
Darwin's and Einstein's correspondence. (a) Cumulative number of letters
sent by Darwin (open circles) and Einstein (solid squares). The solid line
corresponds with a power law growth $N(t)\sim t^{a}$ with $a=2.7$. (b) The
inter-event time distribution associated with the datasets shown in (a).
The solid line represents the power law decay $f(\tau)\sim
\tau^{-\alpha}$, where the exponent $\alpha$ was obtained using
(\ref{alphaaoo}) and the value of $a$ obtained from (a).}

\label{fig1} 

\end{figure}

{\it Regular mail:} In Fig. \ref{fig1}a we plot the cumulative number
letters sent by Darwin and Einstein as a function of time, measured from
the moment the first letter was recorded. In both cases we observe a
growth tendency faster than linear, which is well approximated by the
power law growth $N(t)\sim t^{2.7}$. Since
$N(t)=\int_0^tdt^\prime\lambda(t^\prime)$ this observation corresponds
with a letter sending rate (\ref{lambda}) with $a=3.7$.  Furthermore, both
Darwin and Einstein sent more than 6,000 letters during the time period
considered by this dataset. In this case ($a>1$, $\tau_0\ll T$) we predict
that the inter-event time distribution follows the power law behavior
(\ref{ftauaoo}) with $\alpha\approx2.4\pm0.1$ (\ref{alphaaoo}).  This
prediction is confronted in Fig. \ref{fig1}b with the inter-event time
obtained from the correspondence data, revealing a very good agreement.

\begin{figure}[t]
\centerline{\includegraphics[width=3.8in]{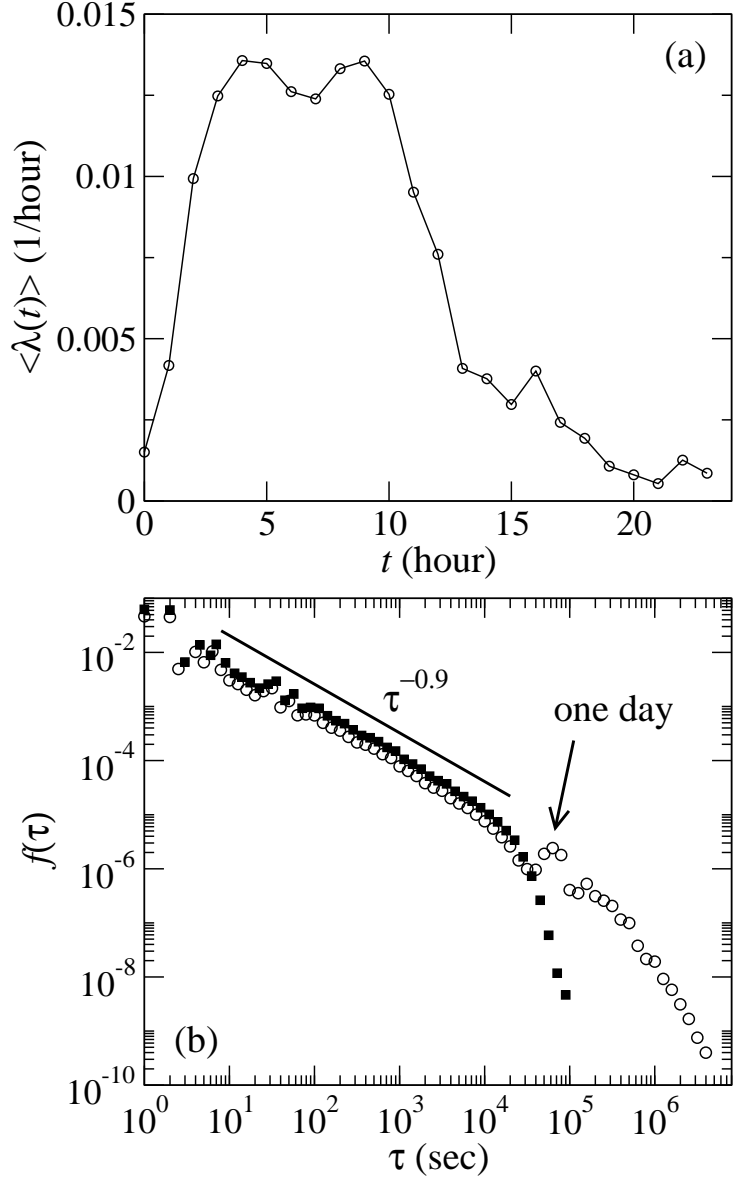}}

\caption{{\bf Email activity:} Statistical properties of the email
activity patterns in an university environment. (a) Email sending rate
average over 81 days and 3,188 users as a function of time. The time was
shift by a constant such that the start of the working hours corresponds
approximately with hour zero. We observe two local maximums associated
with the morning and afternoon peaks of daily activity. More importantly,
this initial relatively high activity is followed by a reducing tendency.
(b)  Aggregated inter-event time distribution of all users. The open
circles are obtained considering both intra-day and inter-day
inter-events, where we can note a local maximum at one day. The solid line
represents the power law decay $f(\tau)\sim \tau^{-\alpha}$ with
$\alpha=0.9$. The solid squares are obtained considering intra-day
inter-events only showing that the power law behavior is determined by
intra-day emails.}

\label{fig2} 
\end{figure}

{\it Email:} Determining the time dependency of $\lambda(t)$ is more
challenging for the email data. If we restrict our analysis to single
users there are only 21 users that sent more than 500 emails. Among them a
few sent more than 1,000 emails but it is questionable how well they
represent the average email user. Therefore, for about 99\% of the users
we do not count with sufficient data to make conclusions about their
individual behavior, being force to analyze their aggregated data.  
Furthermore, email activity patterns are strongly affected by the
circadian rhythm ($T=1$ day) and therefore we can also aggregate data
obtained for different days. In Fig. \ref{fig2}a we plot the email sending
rate averaged over different days and over all users in the dataset as a
function of time. The characteristic features of this plot are: an abrupt
increase following the start of the working hours, two maximums
corresponding with the morning and afternoon activity peaks and a final
decay associated with the end of the working hours.

It is important to note that large inter-event times are associated with
low values of $\lambda$. Therefore, the decrease in the email sending rate
after the working hours determines the tail of the inter-event time
distribution. Based on this we predict that the email activity belongs to
the rate reduction class ($a<1)$.  Furthermore, in average each user sends
an email every two days. In this case ($a>1$, $\tau_0< T$) we predict that
the inter-event time distribution should exhibit a power law behavior
(\ref{ftaua0}) with $0<\alpha<1$ (\ref{alphaa0}). This prediction is
confirmed by the empirical data for the inter-event time distribution (see
Fig. \ref{fig2}b) resulting in $\alpha=0.9\pm0.1$.

\section{Discussion and conclusions}

This work should not be confused with a recent model introduced by
Barab\'asi to characterize the statics of response times
\cite{barabasi05}. The response or waiting time should not be confused
with the inter-event time. For instance, in the context of email activity
the response time is the time interval between the arrival of an email to
our Inbox and the time we answer that particular email. On the other hand,
the inter-event time is the time interval between to consecutive emails
independent of the recipient. For practical applications such as the
design of call centers, web servers, road systems and strategies to halt
epidemic outbreaks the relevant magnitude is the inter-event time.

I have shown that acceleration/reduction tendencies together with some
perception of our past activity rate (\ref{ll}) are sufficient elements to
explain the power law inter-event time distributions observed in two
empirical datasets. Regarding the regular mail correspondence of Darwin
and Einstein the acceleration is probably due to the increase of their
popularity over time. In the case of the email data the rate reduction
could have different origins. We could stop checking emails because we
should do something else or because after we check for new emails the
likelihood that we do it again decreases. The second alternative has a
psychological origin, associated with our expectation that new emails will
not arrive shortly. In practice, the reduction rate of sending emails may
be a combination of these two and factors.

In a more general perspective this work indicates that a minimal model to
characterize human activity patterns is given by two factors: (i) humans
have a perception of their past activity rate and (ii) based on that they
react by accelerating or reducing their activity rate. From the
mathematical point of view memory implies that the progression of the
activity rate is described by integral equations. This is the key element
leading to the power law behavior. These results are relevant to other
human activities where power law inter-event time distributions have been
observed \cite{dezso05,vazquez06}. Before making any general statement,
further research is required to test the validity of the model assumptions
case by case.

\noindent {\bf Acknowledgments:} I thank A.-L. Barab\'asi for helpful 
comments and suggestions and J. G. Oliveira and A.-L. Barab\'asi for 
sharing the Darwin's and Einstein's correspondence data. This work was
supported by NSF ITR 0426737, NSF ACT/SGER 0441089 awards.


\newpage

\end{document}